\documentclass[%
amsmath,amssymb,
prd,
aps,
]{revtex4-2}

\usepackage{graphicx}
\usepackage{dcolumn}
\usepackage{bm}
\usepackage{hyperref}
\usepackage[mathlines]{lineno}

\usepackage{xspace}

\newcommand{\MD}{\ensuremath{M_D}\xspace}
\newcommand{\Mth}{\ensuremath{M_\mathrm{th}}\xspace}
\newcommand{\PBH}{\ensuremath{P_{\,\mathrm{BH}}}\xspace}
\newcommand{\PS}{\ensuremath{P_{\,\mathrm{S}}}\xspace}
\newcommand{\PH}{\ensuremath{\mathcal{P}_{\,\mathrm{H}}}\xspace}
\newcommand{\rH}{\ensuremath{r_{\mathrm{H}}}\xspace}
\newcommand{\qbh}{\textsc{Qbh}\xspace}

\begin{document}


\title{Quantum black holes in the horizon quantum mechanics
model\\ at the Large Hadron Collider}%

\author{Douglas M. Gingrich}
\altaffiliation[Also at ]{TRIUMF, Vancouver, BC V6T 2A3 Canada}
\email{gingrich@ualberta.ca}

\author{Brennan Undseth}
\altaffiliation[Currently at ]{Delft University of
Technology, Netherlands} 
\email{brennan.undseth@gmail.com}

\affiliation{Department of Physics, University of Alberta, Edmonton, AB
T6G 2G7 Canada}

\date{\today}

\begin{abstract}
Quantum black hole production at the Large Hadron Collider is
investigated using the horizon quantum mechanics model.
This model has novel implications for how black holes might be observed
in collider experiments. 
Black hole production is predicted to be possible below the Planck
scale, thus leading to the intriguing possibility that black holes could
be produced even if the Planck scale is slightly above the collider
center of mass energy. 
In addition, the usual anticipated resonance in the black hole mass
distribution is significantly widened in this model.
For values of the Planck scale above the current lower limits, the shape
of the black hole mass distribution is almost independent of the Planck
scale and depends more on the number of extra dimensions.  
These model features suggest the need for alternative search strategies
in collider experiments.
\end{abstract}

\keywords{black holes, extra dimensions, quantum gravity, beyond
Standard Model}
\maketitle

\section{Introduction}

Low-scale gravity provides an interesting possibility for gaining
insight into the hierarchy problem.
A wide variety of models based on different
paradigms~\cite{ArkaniHamed:1998rs,Antoniadis:1998ig,Randall:1999ee}
have been proposed.  
A speculative, but intriguing, possibility of most models is the
production of quantum black holes in hadron
colliders~\cite{Giddings:2001bu,Dimopoulos:2001hw}.

The cross section for black hole production is typically chosen to be
the classical geometric form $\hat{\sigma}\approx\pi r_\mathrm{g}^2$,
where $r_\mathrm{g}$ is the gravitational radius which is a function of
the black hole mass $M$ and depends on the fundamental parameters of the
model. 
In the large extra dimensions paradigm proposed in
Ref.~\cite{ArkaniHamed:1998rs,Antoniadis:1998ig}, the model parameters
are the higher-dimensional Planck scale \MD and total number of
space-time dimensions $D$.
We will consider the case of a tensionless non-rotating spherically
symmetric solution for the gravitational radius~\cite{Myers}.

In proton--proton collisions, only a fraction of the total center of
mass energy $\sqrt{s}$ is available in the hard-scatter process. 
We define $sx_ax_b \equiv s\tau \equiv \hat{s}$, where $x_a$ and $x_b$
are the fractional energies of the two colliding partons (assumed massless)
relative to the proton energies. 
The full particle-level cross section $\sigma$ is obtained from the
parton-level cross section $\hat{\sigma}$ by
using~\cite{Gingrich:2006hm} 

\begin{eqnarray}
\sigma_{pp\to\mathrm{BH+X}}(s) & = & \sum_{a,b} \int_{M^2/s}^1 d\tau
\int_\tau^1 \frac{dx}{x} f_a\left(\frac{\tau}{x}\right) f_b(x)
\Theta(M-\Mth) \hat{\sigma}_\mathrm{ab\to BH} (\hat{s} = M^2)\, ,   
\label{eq1}
\end{eqnarray}

\noindent
where $a$ and $b$ are the parton types in the two protons, and $f_a$ and
$f_b$ are parton distribution functions (PDFs) for the proton.
The sum is over all possible quark and gluon pairings.
While several pre-factors to the cross section have been suggested (see
Ref.~\cite{Gingrich:2006hm} for a summary) they are not important for
this study and will not be considered.

The usual ansatz is that black holes can not be produced with
$M$ below some minimum mass threshold \Mth.
This is emphasized by the use of the Heaviside step function $\Theta$
in Eq.~(\ref{eq1}).  
The value of \Mth is typically taken to be \MD for quantum black holes
or a few times \MD for classical black holes.    
Unfortunately, results depend on the subjective choice of the \Mth
cutoff.

A modification to the typical model of black hole formation in hadron
colliders is made by the horizon quantum mechanics (HQM)
model~\cite{Casadio:2014twa,Casadio:2015jha}. 
The wave function for a localized massive particle (source) is taken to
be a spherically symmetric Gaussian wave packet in $(D-1)$ spatial
dimensions of width $\ell$.
It is postulated that the form of the wave packet in momentum space is
also a Gaussian with width $\Delta = \hbar/\ell$.
The simplest case for black hole formation is considered; a
$D$-dimensional Schwarzschild metric and its classical horizon of radius
$R_D(M)$.
The relativistic mass-shell relation in flat space $E^2 = p^2 + m^2$ is
assumed, where the energy $E$ of the particle is expressed in terms of
the horizon radius $\rH = R_D(E)$ and $m$ is the rest mass of the
source. 
The momentum-space wave function can then be written in terms of the
horizon radius and normalized to give the horizon wave function
$\psi_\mathrm{H}(\rH)$.
The horizon wave function is used to calculate the probability
$\PS(r < \rH)$ that the particle is inside a $(D-1)$-ball of radius \rH
and the probability density $\PH(\rH)$ that the radius of the horizon
equals \rH. 
In this case, the black hole probability depends on the Gaussian width
$\ell$, particle mass $m$, and number of spatial dimensions $D$.
It is further assumed that $\ell = \MD\ell_D/m$ is the Compton
wavelength of the source, which represents the minimum uncertainty in
its size, so that $\Delta = m$, and the probability only depends on $m$
and the number of dimensions $D$.
The system exhibits properties of a black hole when the source is
located within the quantized horizon, with the probability of the system 
being a black hole given by

\begin{equation}
\PBH = \int_0^\infty \PS(r < \rH) \PH(\rH) d\rH\, .
\label{eq2}
\end{equation}

\noindent
Explicit expressions of these probabilities are giving in
Ref.~\cite{Casadio:2014twa,Casadio:2015jha}. 
Qualitatively, the use of the HQM probability in the calculation of the
proton--proton cross section is akin to replacing the step function
located at \Mth with a sigmoid-like function that varies with $M/\MD$
and depends on $D$.

The ingredients that go into deriving Eq.~({\ref{eq2}) are not free of
assumptions. 
In addition, using standard quantum mechanics in the strong gravity
regime is ill defined and the formalism is not free of problems.  
The idea of improving the geometrical cross section by a smoothed step
function is not new~\cite{Mureika:2011hg}.
Using guiding physical principles similar results to Eq.~(\ref{eq2}) can 
be obtained on empirical grounds~\cite{Mureika:2011hg}. 

The common phenomenology of semi-classical microscopic black holes is
not important in this work.
Such objects have significant entropy and Hawking evaporate.
The evaporation process occurs when the mass of the black hole is well
above the Planck scale and thus not close to where HQM effects are
important.  
We thus consider, so called, quantum black holes (QBH), where the object
has an event horizon but negligible entropy, and behaves more like a
particle in its decay to a few-body -- two in our case -- final state.   
Such objects by definition have mass close to the Planck scale and are
significantly affected by the HQM model.

The purpose of the work presented here is to evaluate the impact of the
HQM model on the production of quantum black holes with emphasis on the
signatures for experiments at the Large Hadron Collider (LHC). 
We begin with a brief description of Monte Carlo (MC) event generation
in the HQM model, with more details of the implementation described in
Appendix~\ref{appA}. 
We discuss the effects of HQM on the total proton--proton cross section
and the differential proton--proton cross section as a function of $M$. 
The possibility of quantum black hole detection in the HQM model in LHC
experiments is discussed. 
A previous publication~\cite{Arsene:2016kvf} on this topic made use of
ATLAS and CMS results from about 20~fb$^{-1}$ of data at $\sqrt{s} =
8$~TeV. 

We make use of the following conventions.
When comparing models, the QBH model refers to the quantum black hole
model with Heaviside step function turn-on typically used by ATLAS and
CMS searches at $\sqrt{s}$ of 
7~TeV~\cite{Aad:2011aj,ATLAS:2012pu,CMS:2012yf}, 
8~TeV~\cite{Aad:2013cva,Aad:2013gma,Aad:2014cka,Aad:2014aqa,Khachatryan:2015sja,Khachatryan:2016ovq},
and
13~TeV~\cite{ATLAS:2015nsi,Aaboud:2016hmk,Aaboud:2017yvp,Sirunyan:2017ygf,Aaboud:2017nak,Sirunyan:2018zhy,Aaboud:2018jff,Aad:2019hjw}
that does not include any HQM effects.  
The HQM model will be the model with horizon quantum mechanics effects
included. 
The only difference between these two models is their production turn-on
behaviour in $M/\MD$ for different $D$.  
The total number of space-time dimensions $D = n + 4$, where $n$ is the
number of extra dimensions.

\section{Black hole production probability}

For the purpose of cross section calculations along with event
generation, the \qbh 3.00 MC quantum black hole event
generator\footnote{We use QBH to refer to the quantum black hole model
and \qbh to refer to the quantum black hole generator of the same name.}
is used~\cite{Gingrich:2009da}.  
In this model~\cite{Meade:2007sz,Calmet:2008dg,Gingrich:2009hj}, we
consider tensionless non-rotating black holes. 
The generator only allows for dominant two-body decay of the QBH states.
The leading-order CTEQ6L1~\cite{Pumplin:2002vw} PDF set is used for the
hard-scattering process. 
Considering only two-body decays and using the CTEQ6L1 PDF set are
consistent with the ATLAS and CMS experiment's QBH searches. 
The default settings in \qbh are used and the proton--proton center of
mass energy is set to 13~TeV. 
The only parameters that are varied are \MD and $D$.
Cross section calculations in \qbh are independent of the number of
events generated.
For kinematic distributions, 21000 events were generated for each
$(D,\MD)$ pair.
We work at the parton level and do not hadronize the partons or decay
the final state particles; a hadron is considered as a single jet.
No energy-momentum smearing or detector simulation has been performed. 
HQM effects are added to the proton--proton cross section by including
the factor \PBH of Eq.~(\ref{eq2}) into Eq.~(\ref{eq1}):  

\begin{eqnarray}
\sigma_{pp\to\mathrm{BH+X}}(s) & = & \sum_{a,b} \int_{M^2/s}^1 d\tau
\int_\tau^1 \frac{dx}{x} f_a\left(\frac{\tau}{x}\right)
f_b(x)\PBH(M)\hat{\sigma}_\mathrm{ab\to BH} (\hat{s} = M^2)\, ,   
\label{eq3}
\end{eqnarray}

\noindent
where \PBH requires another numerical integration.
The cross section formula is now independent of \Mth and the model has
one less free parameter.

In order to visualise how the HQM probability varies with $M, \MD$, and
$D$, we have computed the integral in Eq.~(\ref{eq2}) explicitly, as
shown in Fig.~\ref{fig01}. 

\begin{figure}[tb]
\centering
\includegraphics[width=\columnwidth]{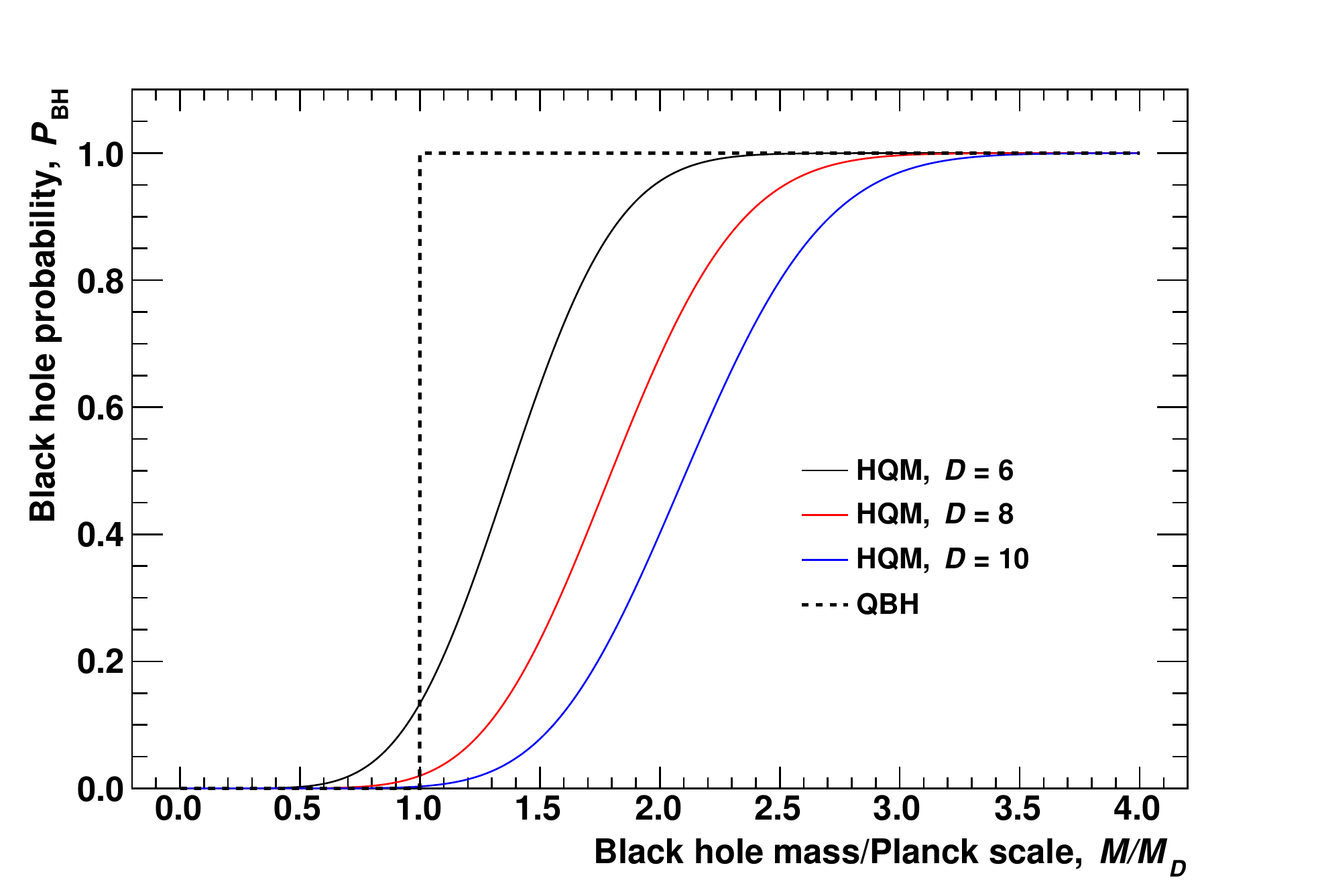}
\caption{\label{fig01}Horizon quantum mechanics (HQM) probability curves
\PBH versus black hole mass $M$ relative to the Planck scale \MD for
selected total number of space-time dimensions $D$.    
The dashed black line represents the step function used in quantum black
hole (QBH) models.} 
\end{figure}

The probability curves suggest some interesting phenomena that are not
seen in the QBH model.
First, instead of a step function at $M = \MD$, the new curves are
smooth. 
The most notable consequence is that there is a finite probability that
a black hole can be formed with $M < \MD$.
Second, we see that the probability for a black hole to be produced near 
\MD is suppressed for high $D$.
In other words, one generally expects more black holes to be produced
for low $D$.
This is at odds with the usual effect of dimensionality in the QBH
model, where greater $D$ corresponds to a greater geometric cross
section. 
A third observation is that most of the curve is significantly above
the value of $M/\MD = 1$.
And lastly, the slope in the curves at $\PBH = 0.5$ are not particularly
steep. 

We can roughly quantify the extent to which the \PBH curves create a
threshold in the $M$ distribution by considering the midpoint of
each curve as the point where $\PBH = 0.5$. 
These values are shown in Table~\ref{tab1}.
For $D=6$, the black hole mass threshold rises to slightly above
the usual \MD threshold in the QBH model. 
For $D=10$, the threshold is more than twice \MD.
This means that more dimensions will cause heavy suppression of black 
hole production in the HQM model, unlike the QBH model in which more
black holes will be produced at higher $D$.
The actual values in Table~\ref{tab1} are model dependent but the trends
are indicative. 

\begin{table}[htb]
\caption{\label{tab1}Ratio of black hole mass $M$ to Planck scale \MD at
  $\PBH = 0.5$ for different total number of space-time dimensions $D$
  in the horizon quantum mechanics model.} 
\begin{ruledtabular}
\begin{tabular}{ccccccc}
$D$ & 6 & 7 & 8 & 9 & 10 & 11\\
$M/\MD$ & 1.4 & 1.6 & 1.8 & 2.0 & 2.1 & 2.2\\
\end{tabular}
\end{ruledtabular}
\end{table}

\section{Proton--proton total cross section}

We start by analyzing how the inclusion of HQM impacts the 
proton--proton total cross section as a function of \MD and $D$.  
There are two competing factors at play.
On one hand, we are multiplying the parton-level cross section by a
factor between 0 and 1, which in general decreases the cross section. 
On the other hand, we are considering a wider range of possible $M$
than in the QBH model. 
In addition, while it is unreasonable to think of producing events in
the QBH model if $\MD > \sqrt{s}$, the smooth cutoff imposed by HQM
allows for black holes when \MD is above the collider energy.  
The phenomena are shown in Fig.~\ref{fig02}.

\begin{figure}[tb]
\centering
\includegraphics[width=\columnwidth]{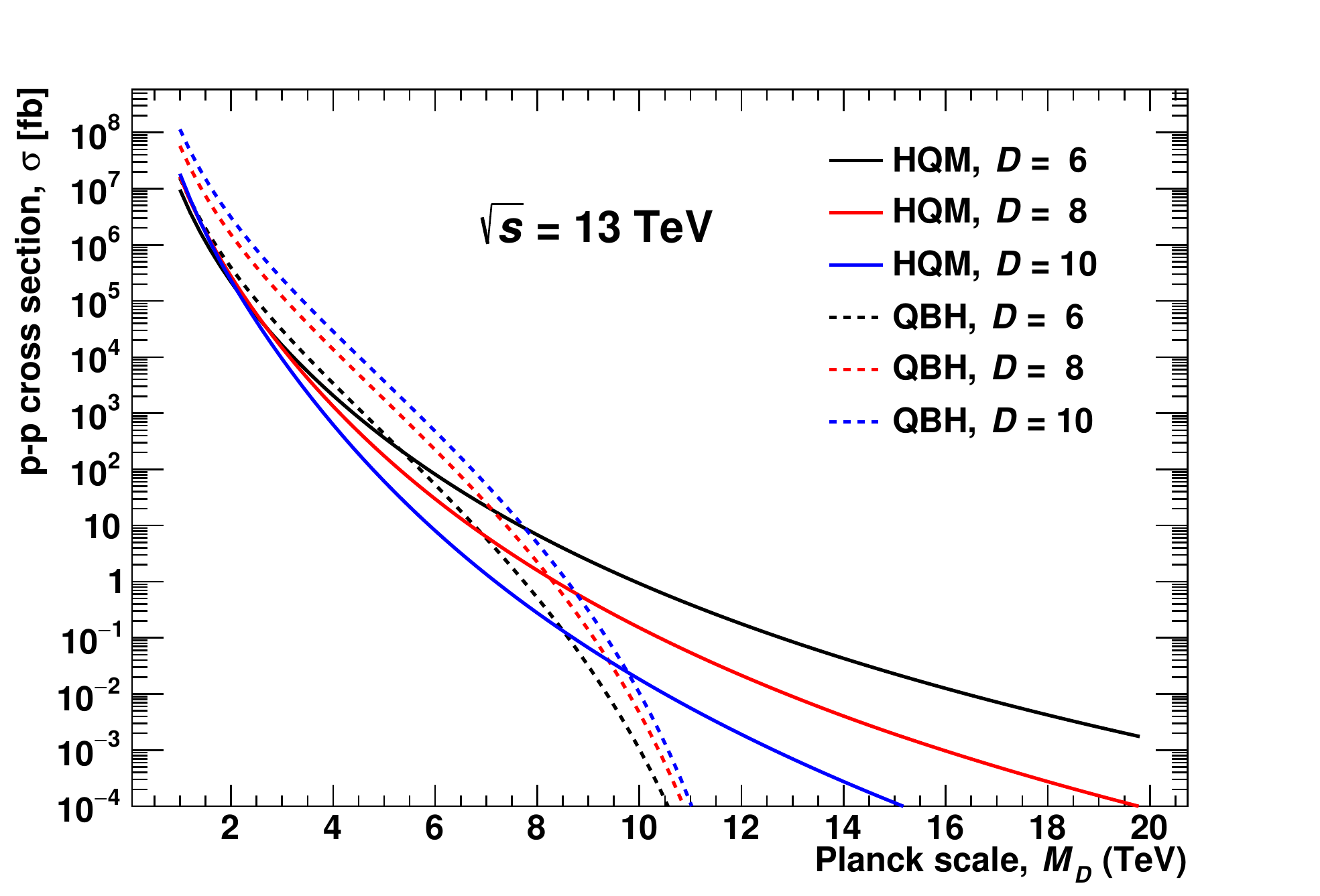}
\caption{\label{fig02}Proton--proton total cross section $\sigma$ versus
Planck scale \MD at a center of mass energy of 13~TeV.
Curves for different models and total number of space-time dimensions
$D$ are shown.
Solid curves are used for the horizon quantum mechanics (HQM) model and
dashed curves are used for the quantum black hole (QBH) model.}  
\end{figure}

The inclusion of HQM suppresses the total cross section for low
\MD but predicts a higher cross section than the QBH model at high
\MD. 
It is also interesting to note how the role of dimensionality is
reversed in the two models.
For a given \MD, higher cross sections occur at lower $D$ in the
HQM model, except for a small region below about 2~TeV.
Also, in the HQM model the cross section at a given \MD is significantly
different for different $D$ as \MD increase. 
Thus over most of the \MD range, dimensionality is significantly more
important in the HQM model. 

It is also useful to determine the \MD value at which the HQM model cross
section crosses over the QBH model cross section, and thus where the HQM
model might become more significant.
For $D=6$, $D=8$, and $D=10$, the crossovers in \MD occur at
approximately 5.4~TeV, 8.2~TeV, and 9.7~TeV, respectively. 
To understand which region of \MD is interesting, we consider the
current lower-limits, at the 95\% confidence level, on \MD of 9.9~TeV,
6.3~TeV, and 5.3~TeV for $D=6$, $D=8$, and $D=10$, respectively, set by
the CMS~\cite{Sirunyan:2017jix} and ATLAS~\cite{Aaboud:2017phn}
experiments. 
At these \MD limits, black hole production in the HQM model is still
well below the QBH model except for $D=6$ where the HQM model predicts a
cross section of about three orders of magnitude higher than the QBH
model.  

The lower limits on \MD are based on graviton searches in
the same large extra dimensions
paradigm~\cite{ArkaniHamed:1998rs,Antoniadis:1998ig} as used for black 
hole models, and we thus take them to be applicable to both the QBH and
HQM models considered here.  
Searches for QBHs have set limits on \Mth (or \MD as a function of
\Mth), and thus do not constrain the HQM model; there are currently no
limits on \MD using the HQM model. 

The most glaring difference between models occur above the \MD lower
limits. 
While the QBH model cross sections falls sharply as $\Mth = \MD$ is
pushed toward $\sqrt{s}$, the HQM model cross sections exhibit a more
gradual drop that becomes less steep at higher \MD. 
This results in some notable properties unique to the HQM model.
First, black holes may be produced even if $\MD > \sqrt{s}$.  
Second, since the cross sections do not converge to zero at $\MD =
\sqrt{s}$, dimensionality plays a greater role at high \MD.

\begin{figure}[htb]
\centering
\includegraphics[width=\columnwidth]{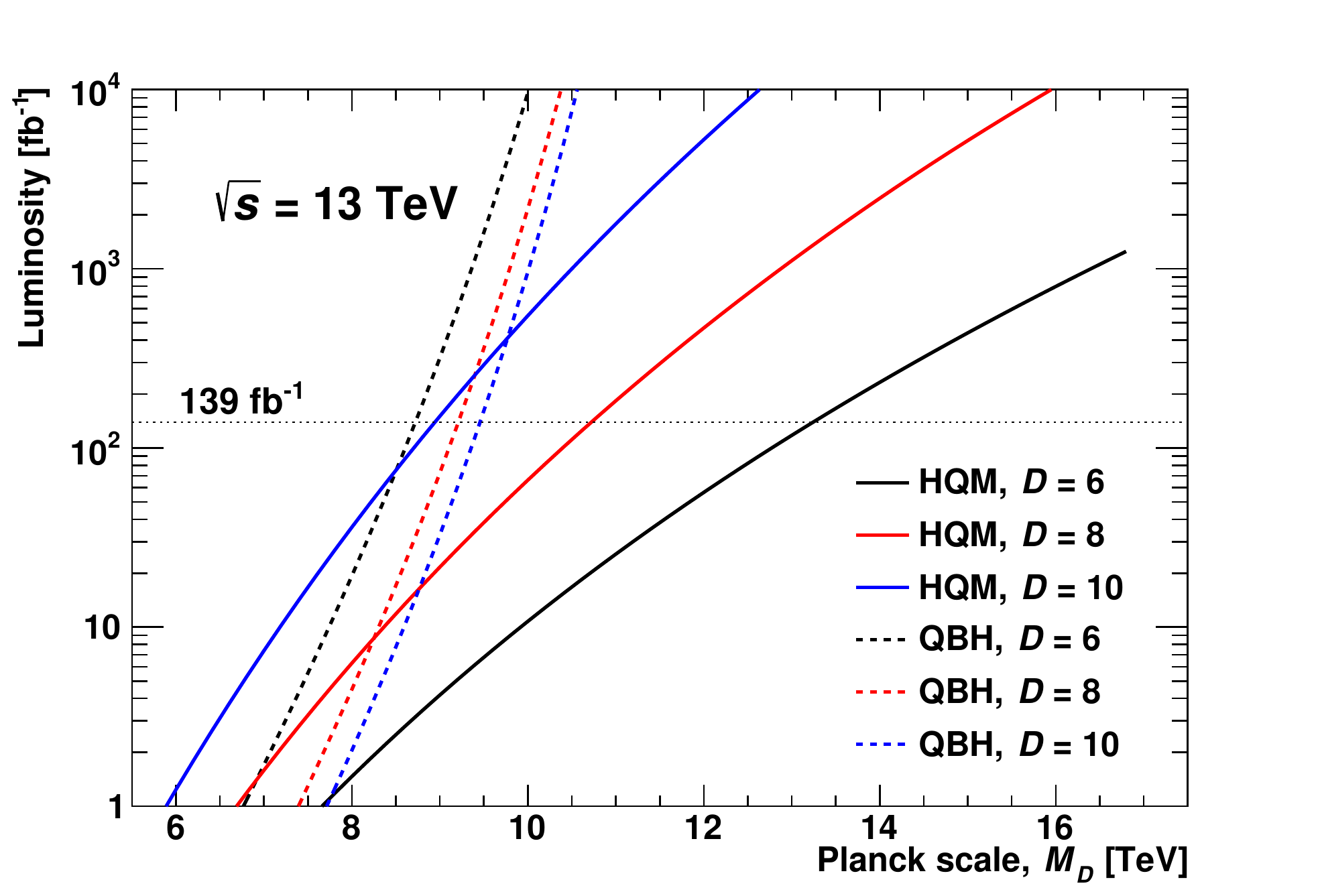}
\caption{\label{fig03}Luminosity required to produce ten black hole
events as a function of Planck scale \MD at a center of mass energy
of 13~TeV.
Curves for different models and total number of space-time dimensions
$D$ are shown.
Solid curves are used for the horizon quantum mechanics (HQM) model and
dashed curves are used for the quantum black hole (QBH) model.  
The horizontal dotted line represents a luminosity of 139~fb$^{-1}$.}  
\end{figure}

Of particular importance for observing quantum black holes in experiments
is the number of black hole events we are able to produce. 
Typically, a minimum of ten signal events is sought to form a reasonable
claim of discovery\footnote{At this point, we are assuming a perfect
search for black holes.}
In Fig.~\ref{fig03}, we plot the luminosity required to produce ten
events in proton--proton collisions at $\sqrt{s} = 13$~TeV.  
Analysis performed by ATLAS and CMS using the full run-2 dataset
typically quote a luminosity of about 139~fb$^{-1}$. 
Using this luminosity, more than ten events can be produced in the QBH
model for \MD less than about 8.7~TeV, 9.2~TeV, and 9.5~TeV for $D =
6$, $D = 8$, and $D = 10$, respectively. 
The lower limits on \MD would exclude $D = 6$ black holes in the QBH
model.
The current best lower limit from a direct QBH search is $\Mth = \MD >
9.4$~TeV for $D = 10$~\cite{Aad:2019hjw}.
Even with a luminosity of 1~ab$^{-1}$ at $\sqrt{s} = 13$~TeV, the limit on
\Mth in the QBH model is unlikely to go above about 10.5~TeV. 
Thus, the QBH model is being significantly restricted even at current
luminosities.

The LHC is able to produce black holes at much higher values of \MD in
the HQM model for most $D$. 
At a current luminosity of 139~fb$^{-1}$, values of \MD in the HQM
model are not constrained by the lower limits on \MD, and quantum
black holes could exist in the LHC experiment's current datasets.
However, as we will see next it will be non-trivial to detect HQM
black holes in current ATLAS and CMS datasets even if produced.

\section{Proton--proton differential cross section}

The inclusion of HQM in quantum black hole production has notable
implications on the $M$ distribution of black holes. 
Since the cross sections of QBH and HQM models typically differ by
over an order of magnitude (except at very low \MD and near the crossing), 
it is illustrative to compare the normalized shapes of distributions for
\MD of interest.
Figure~\ref{fig04} compares $M$ distributions for four selected values
of \MD and $D=10$.

For a small \MD, the HQM model gives the peak structure of the QBH
model, but this changes for higher \MD, and $M$ is distributed over a
wide range: $2 \lesssim M \lesssim 10$~TeV.
This difference in shape is a direct consequence of the shapes of the
PDFs and the \PBH curve from HQM. 
The PDFs fall rapidly as parton energies approach $\sqrt{s}/2$.
For $\MD = 12$~TeV in the QBH model, a very small cross section is
expected since $M$ is limited to the range $12 < M < 13$~TeV.  
In the $\MD = 12$~TeV HQM model, the lower mass for black holes is
dictated by the \PBH curve.  
Black hole masses below 2~TeV are suppressed since
$\PBH\approx 0$, and likewise black holes with mass above about 10~TeV
are suppressed by the PDFs. 
This interplay in the HQM model between the convolution of PDFs and \PBH
gives rise to the shape of the $M$ distributions.

\begin{figure*}[htb]
\centering
\includegraphics[width=7.2cm]{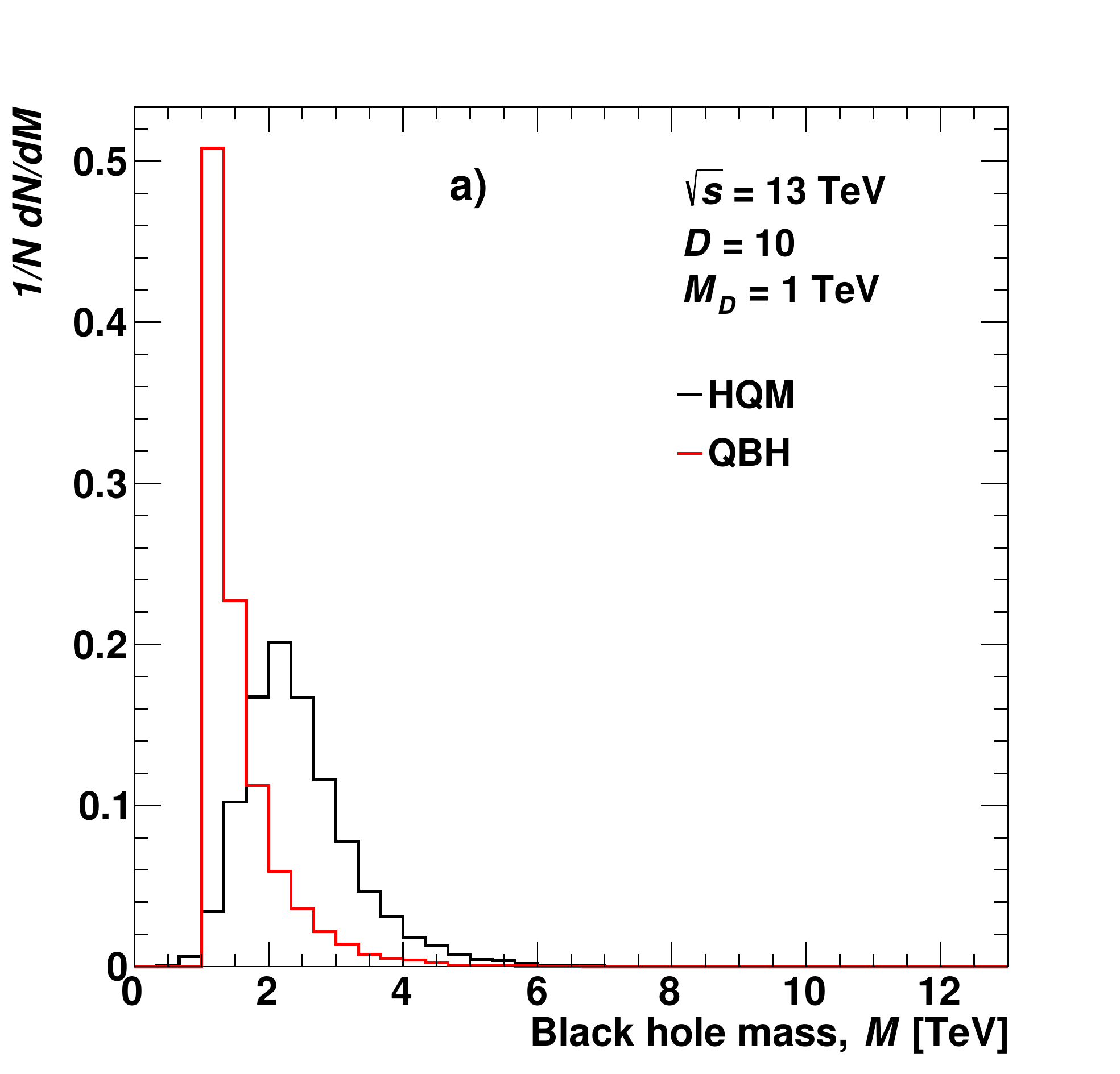}
\includegraphics[width=7.2cm]{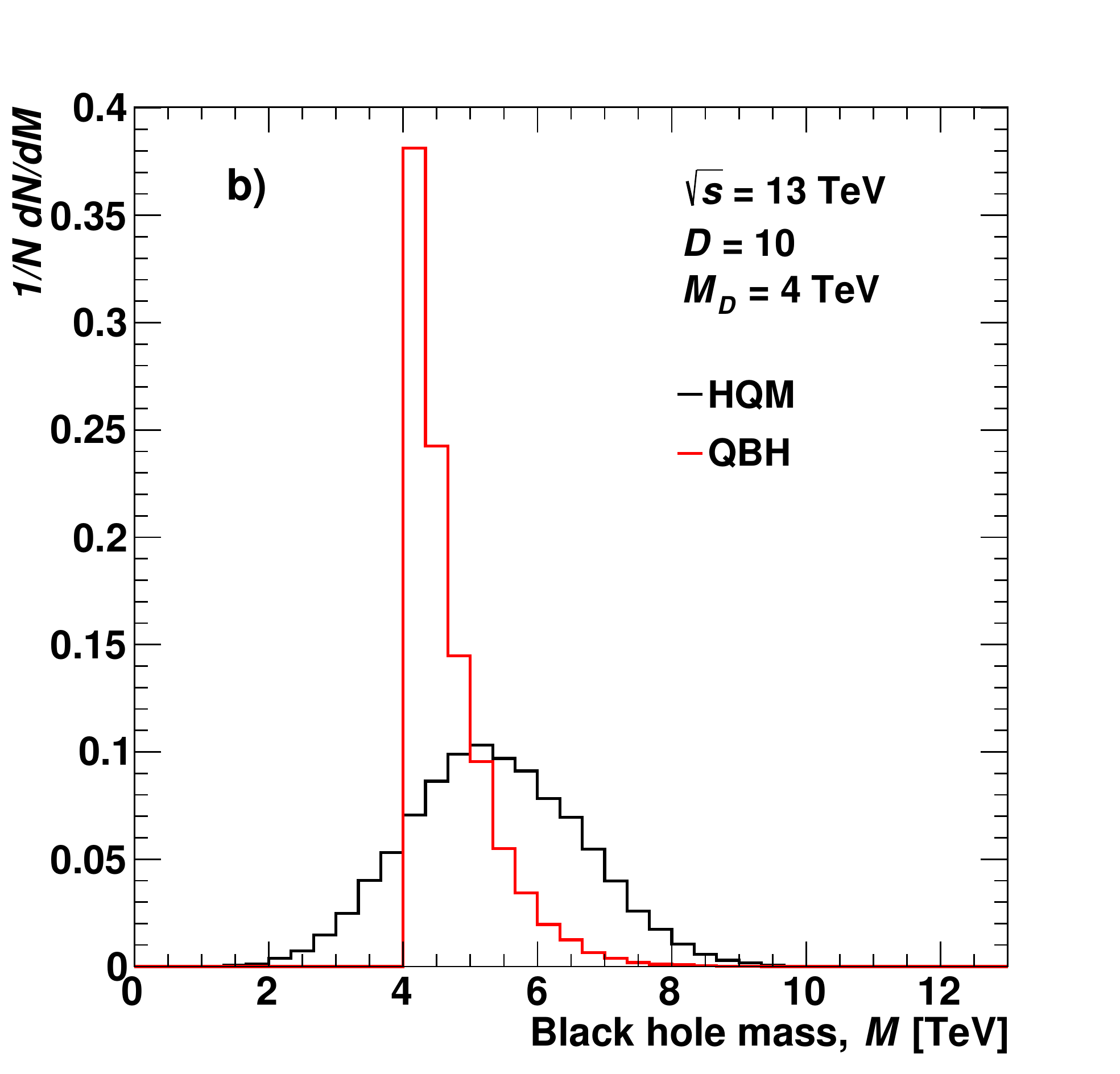}
\includegraphics[width=7.2cm]{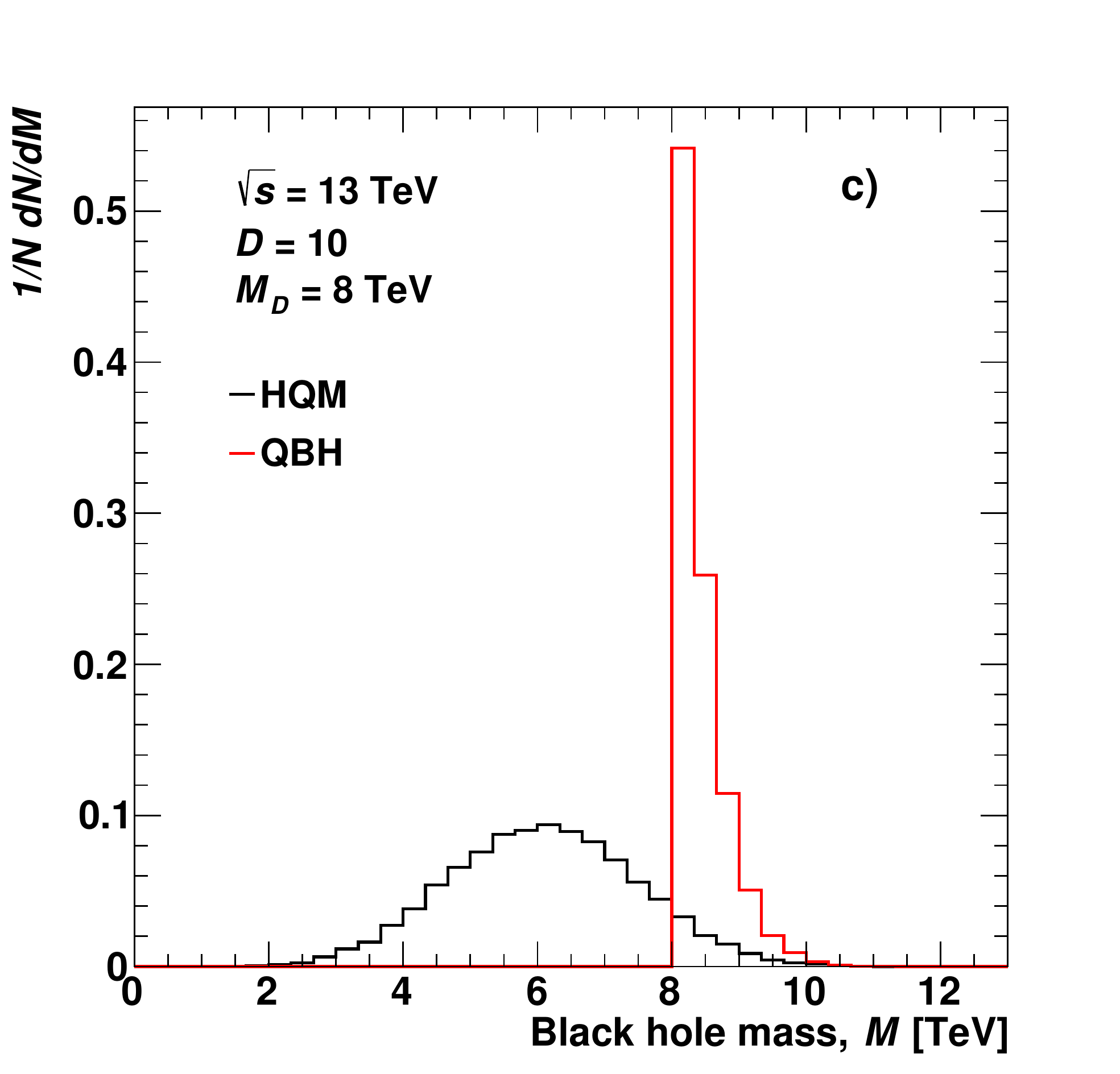}
\includegraphics[width=7.2cm]{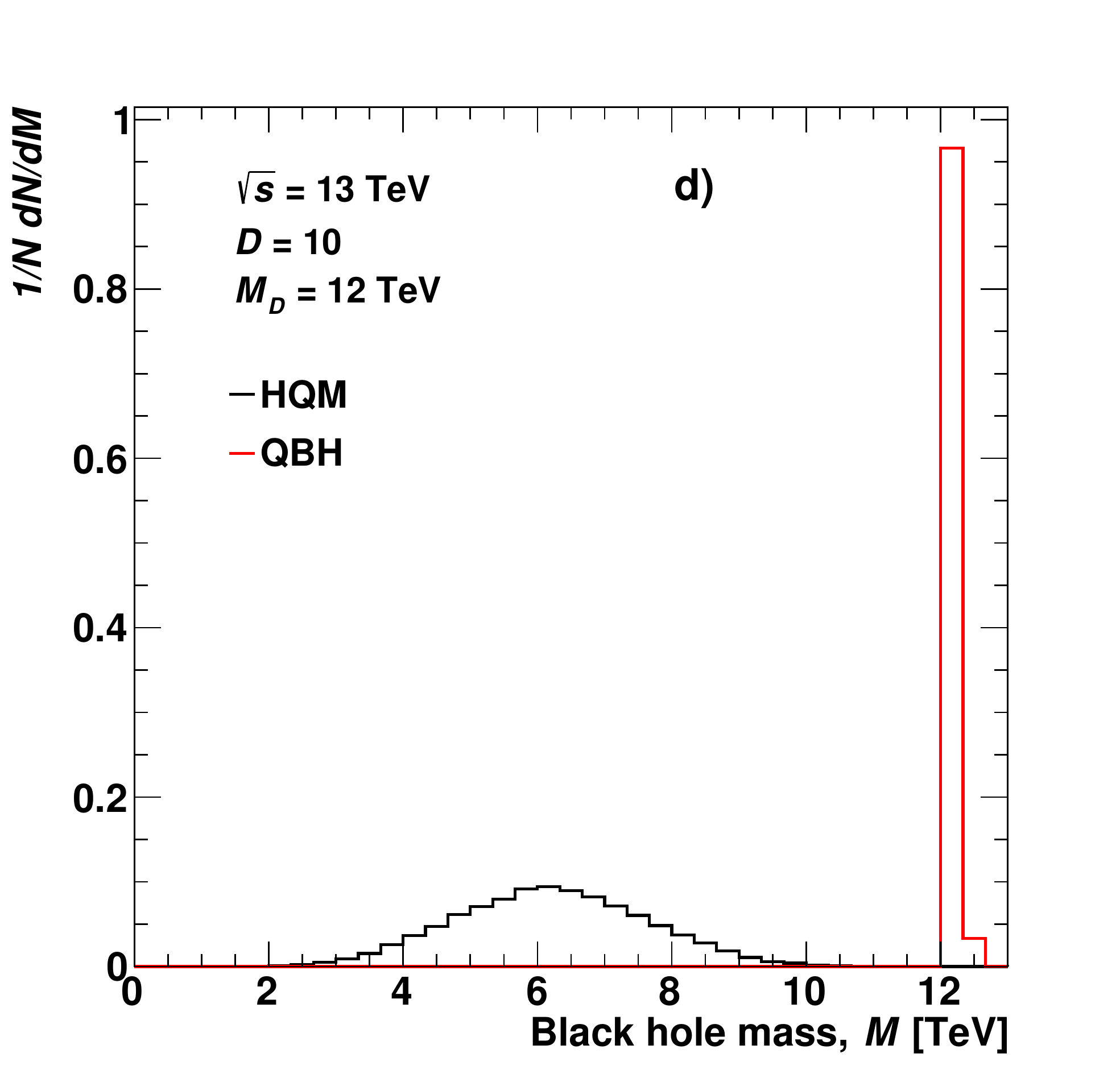}
\caption{\label{fig04}Quantum black hole (QBH) model and horizon quantum 
mechanics (HQM) model mass $M$ distributions normalized to unity for a)
$\MD = 1~$TeV, b) $\MD = 4$~TeV, c) $\MD = 8$~TeV, and d) $\MD =
12$~TeV. 
The center of mass energy is 13~TeV and $D=10$.}
\end{figure*}

The peak in the QBH $M$ distribution moves up with increasing \MD since
the model's definition of \Mth is a strict cutoff in $M$. 
In contrast, the HQM model $M$ distribution does not appear to shift up
much above $\MD \gtrsim 7$~TeV.
This phenomena is explored further in Fig.~\ref{fig05}. 
While the QBH model $M$ distribution moves up with increasing \MD acting
as a minimum mass threshold, the HQM model $M$ distributions are much
more spread out and the shape of the distributions do not change
significantly once \MD exceeds a few TeV.
We also observe that in the HQM model it is very difficult to produce
black hole masses above $\sim 11$~TeV, even though \MD is not limited.

\begin{figure}[tbp]
\centering
\includegraphics[width=0.9\columnwidth]{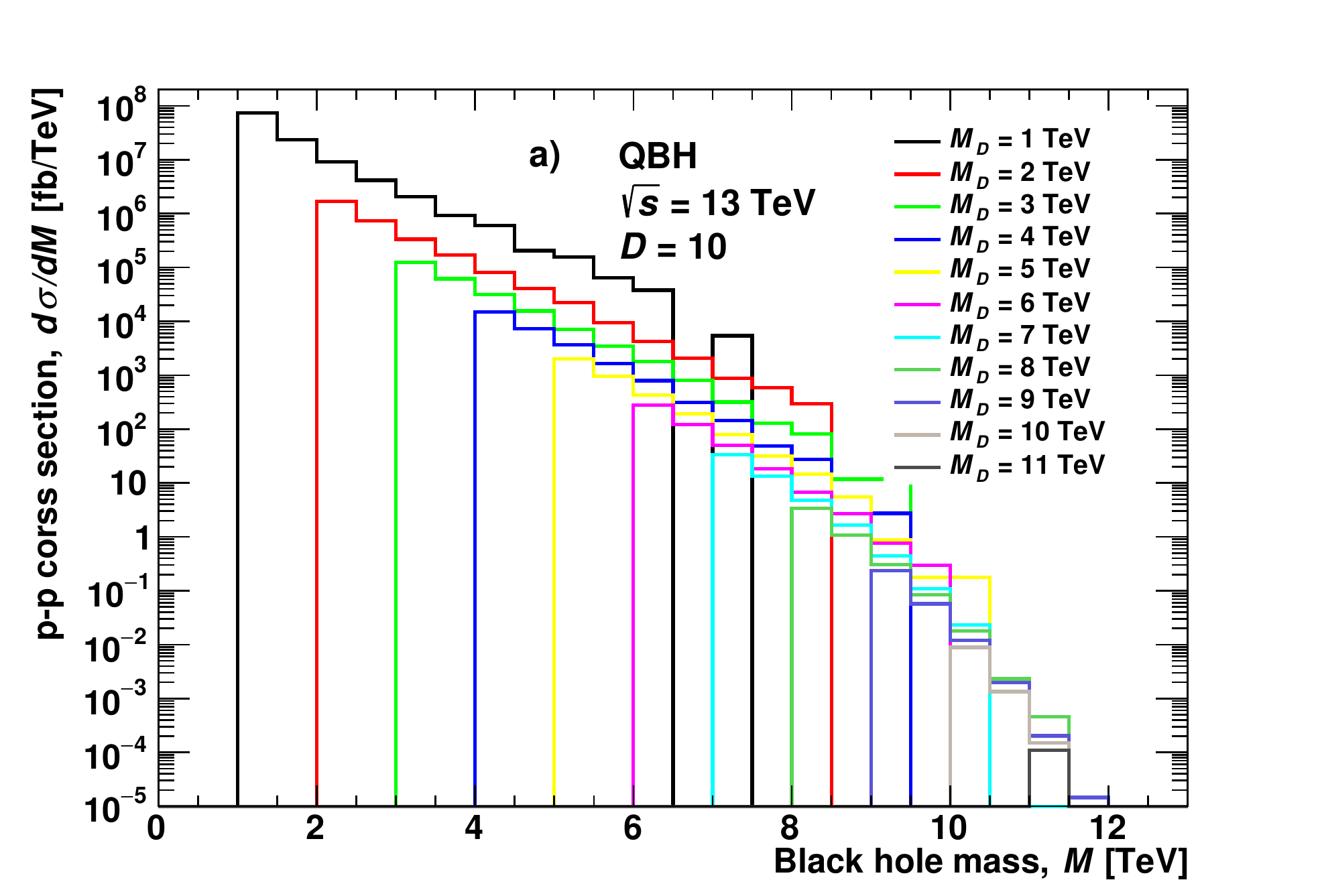}
\includegraphics[width=0.9\columnwidth]{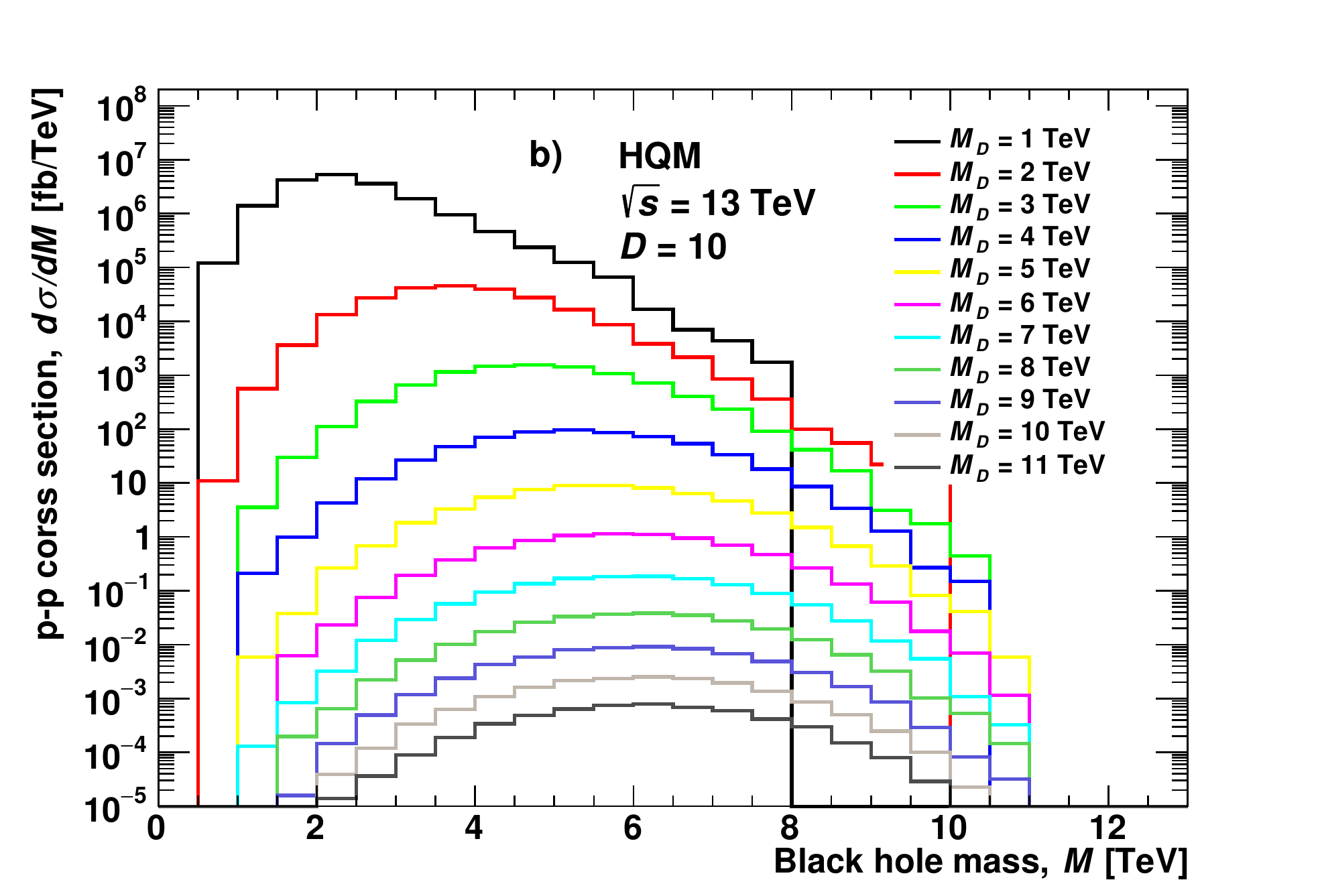}
\caption{\label{fig05}Quantum black hole (QBH) model a) and horizon
quantum mechanics (HQM) model b) proton--proton differential cross
sections $d\sigma/dM$ versus black hole mass $M$ for selected values of
the Planck scale \MD.  
The center of mass energy is 13~TeV and total number of space-time
dimensions $D=10$.}
\end{figure}

The progression of black hole $M$ distributions with \MD in both models
is shown in Fig.~\ref{fig06}, which plots the mean $M$ as a function of 
\MD.  
The QBH model curve gives exactly what is expected, since most black
holes are produced with mass \MD, a linear increase in the mean $M$ is 
observed for all $D$.   
This is in contrast to the HQM model which resembles a linear increase
only for small \MD and then levels off at a constant mean $M$ for
$\MD \gtrsim 8$~TeV. 
The value of the mean $M$ to which the trend converges is dependent on
$D$. 
The reason for this is an interplay between the \PBH curves
which approach zero as $M$ approaches zero and the PDFs
which approach zero as $M$ approaches $\sqrt{s}$. 
The consequence is a ``pinching off'' that serves to create a mass
distribution that does not change shape significantly between the two
mass regions where the production of black holes is vanishingly small. 
The mean $M$ increases with $D$ due to the \PBH curve being shifting
higher in $M/\MD$ with increasing $D$, as previously shown in
Fig.~\ref{fig01}. 

\begin{figure}[htb]
\centering
\includegraphics[width=\columnwidth]{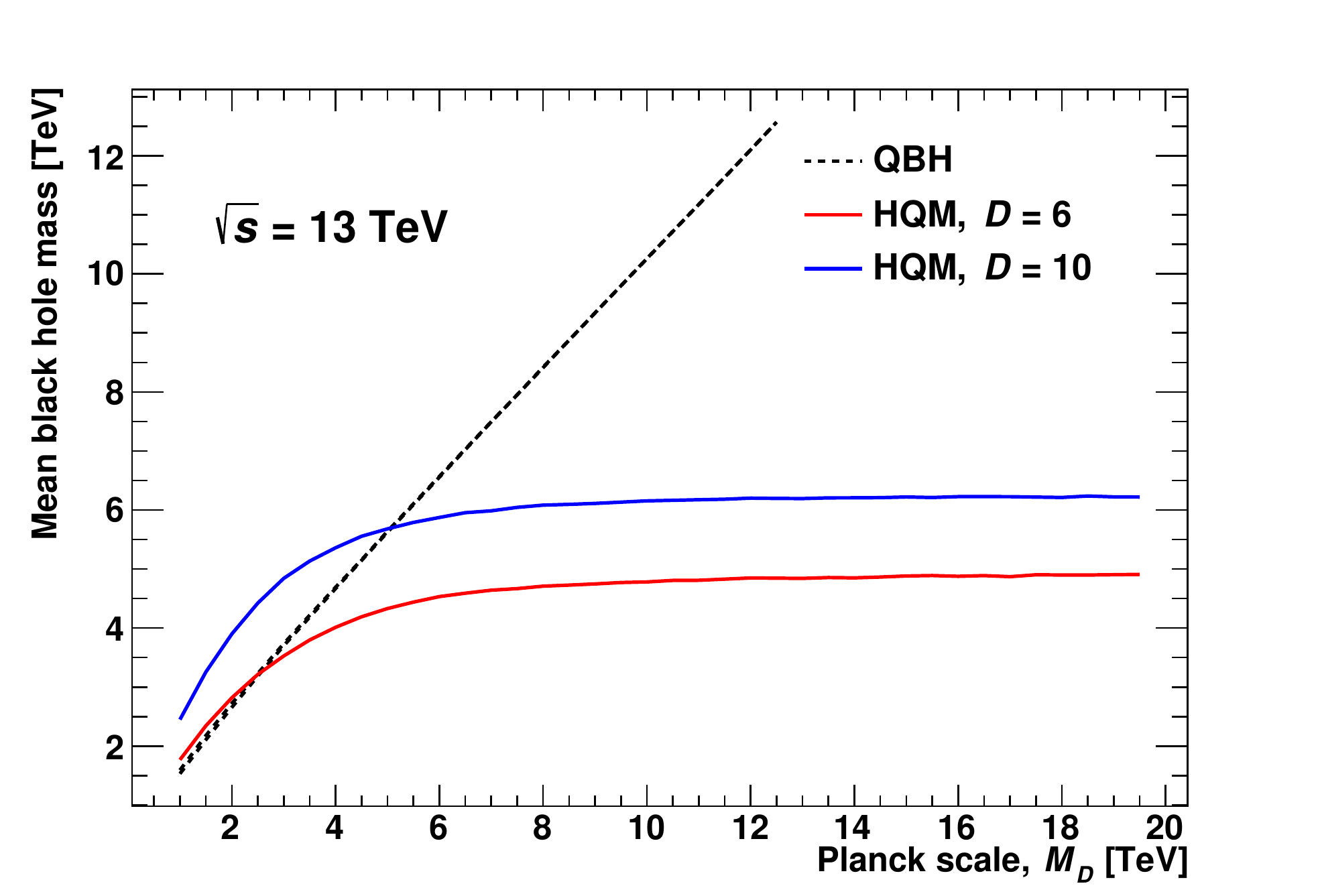}
\caption{\label{fig06}Mean mass of black hole events as a
function of Planck scale \MD at a center of mass energy of 13~TeV.
Quantum black hole (QBH) and the horizon quantum mechanics (HQM) models
are shown for two values of total number of space-time dimensions $D$.}
\end{figure}

Finally, the shape of the HQM model $M$ distribution has implications on
how black holes in this model may be detected in the ATLAS and CMS
experiments. 
In the QBH model, black holes are expected to predominantly decay into
two-body final states. 
The majority of these decay products would be quarks and gluons that
would hadronize to produce jets. 
For this reason, ATLAS and CMS have searched for resonances in the
mass distribution of dijet events. 
The branching fraction to dijets is greater than 96\%~\cite{Aaboud:2017yvp}.
The experiments have taken the branching fraction to dijets to be unity
and have accounted for events with less than two jets in the efficiency.
In our study, we have ignored this inefficiency.

To investigate how black holes in the HQM model would appear in 
these searches, we use 139~fb$^{-1}$ of ATLAS data recorded during run-2
at $\sqrt{s} = 13$~TeV~\cite{Aad:2019hjw}~\footnote{The data is taken
from the HEPData repository
\url{https://www.hepdata.net/record/ins1759712}.}.   
Quantum black hole events are simulated using the same selection criteria,
at the particle level, as in the ATLAS analysis.
We understand that particle-level selection will only roughly emulate
the geometrical acceptance of events in the the ATLAS detector, but the
signal yields should be indicative of a full experimental
analysis.\footnote{The quantum black holes we consider are only
effected by the rapidity requirements; the simulated events pass the
transverse momentum, invariant mass, and azimuthal angle requirements.}    

In Fig.~\ref{fig07}, the black hole events have been scale by the cross
section times luminosity divided by the number of generated events.
Fractional events are possible.
Figure~\ref{fig07}a) shows an example QBH resonance for $\MD = 9.5$~TeV and
$D = 10$.
This resonance is beyond the highest dijet mass event obtained by ATLAS.
In addition, the decisive lack of such a resonance structure in the
dijet mass spectrum has allowed ATLAS to limit black holes in the QBH
model to $\Mth > 9.4$~TeV for $D=10$ at the 95\% confidence
level~\cite{Aad:2019hjw}. 
Thus, the QBH model in its simplest form is close to being ruled out.

\begin{figure*}[htb]
\centering
\includegraphics[width=7cm]{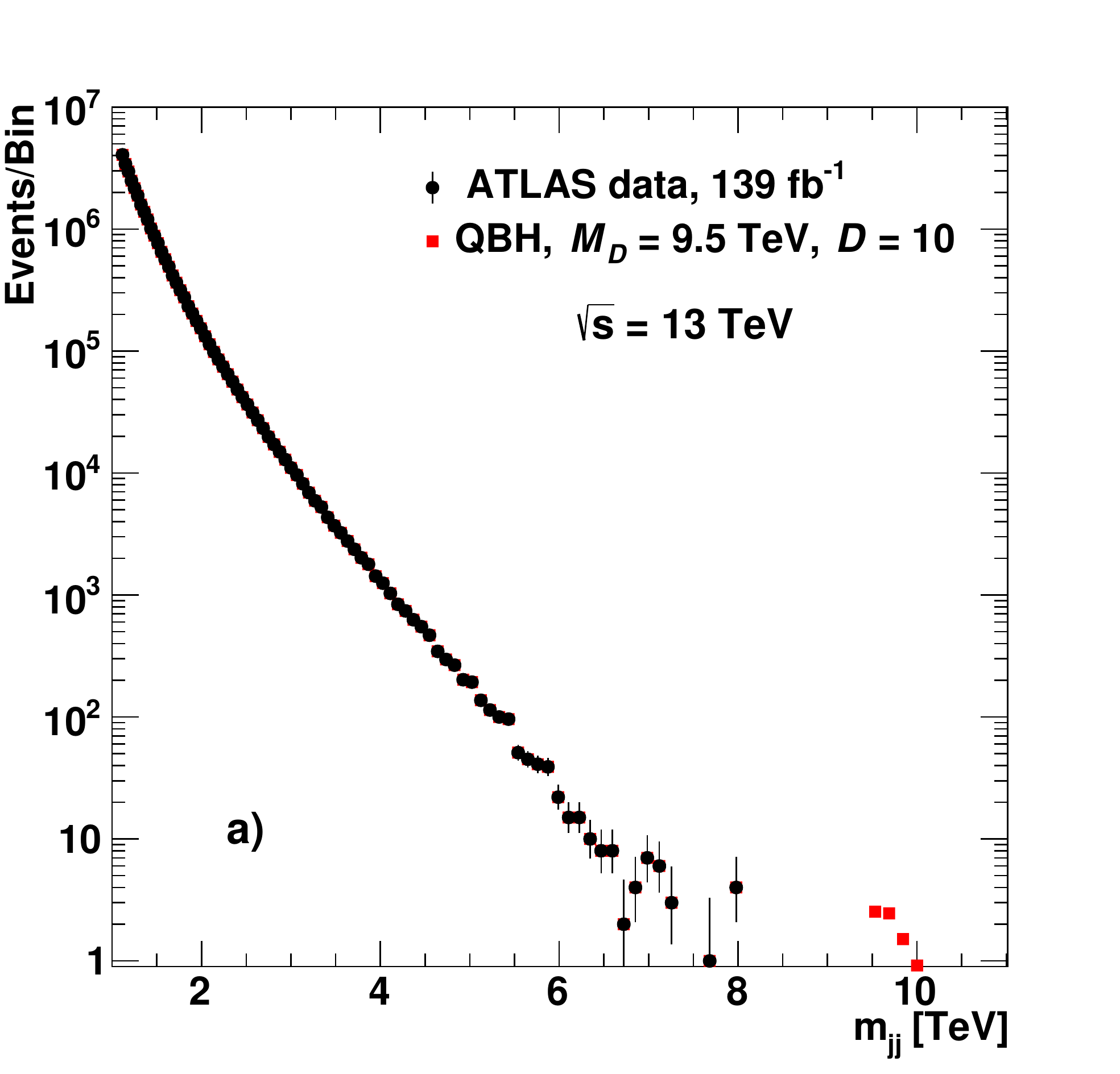}
\includegraphics[width=7cm]{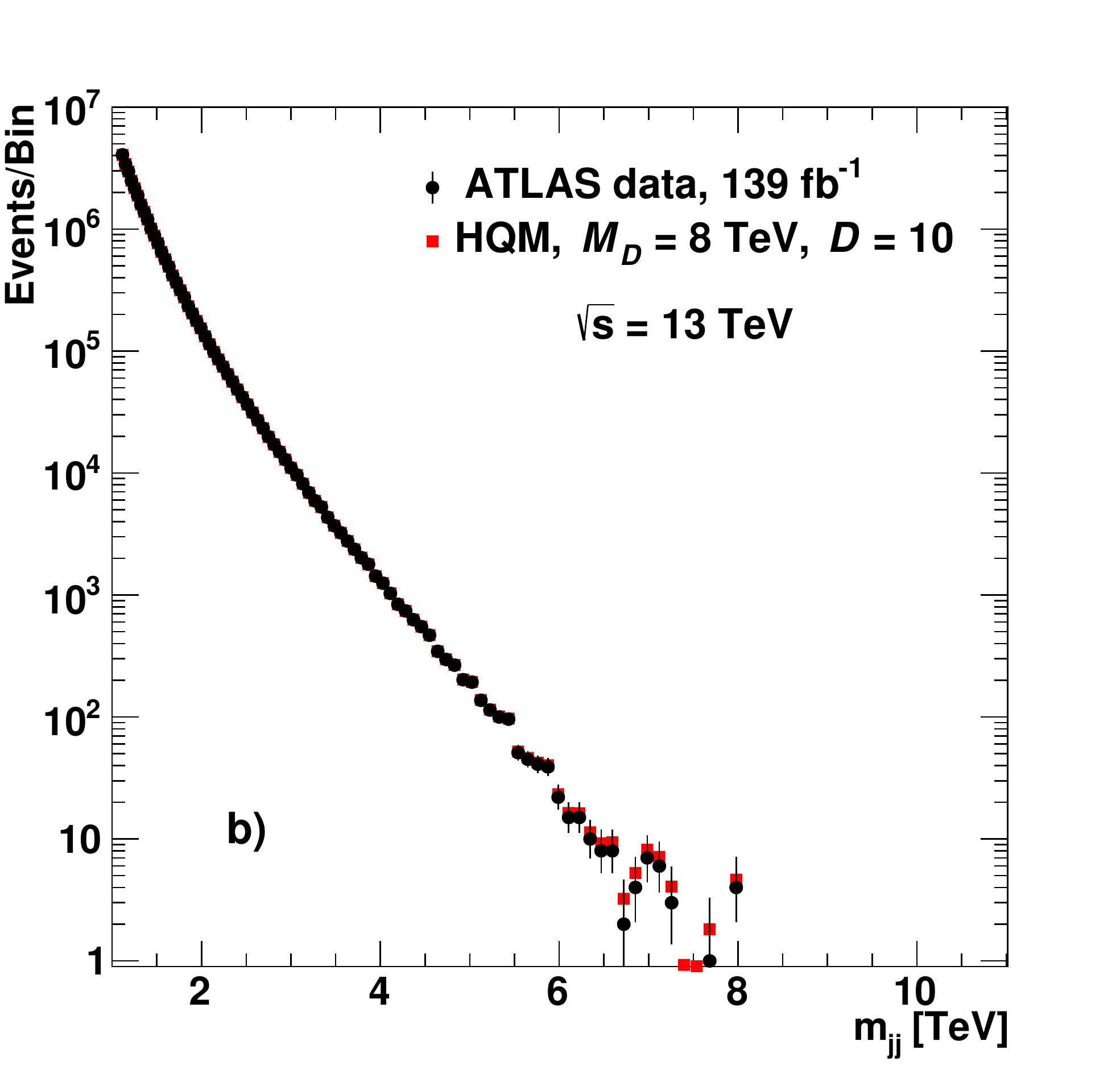}
\includegraphics[width=7cm]{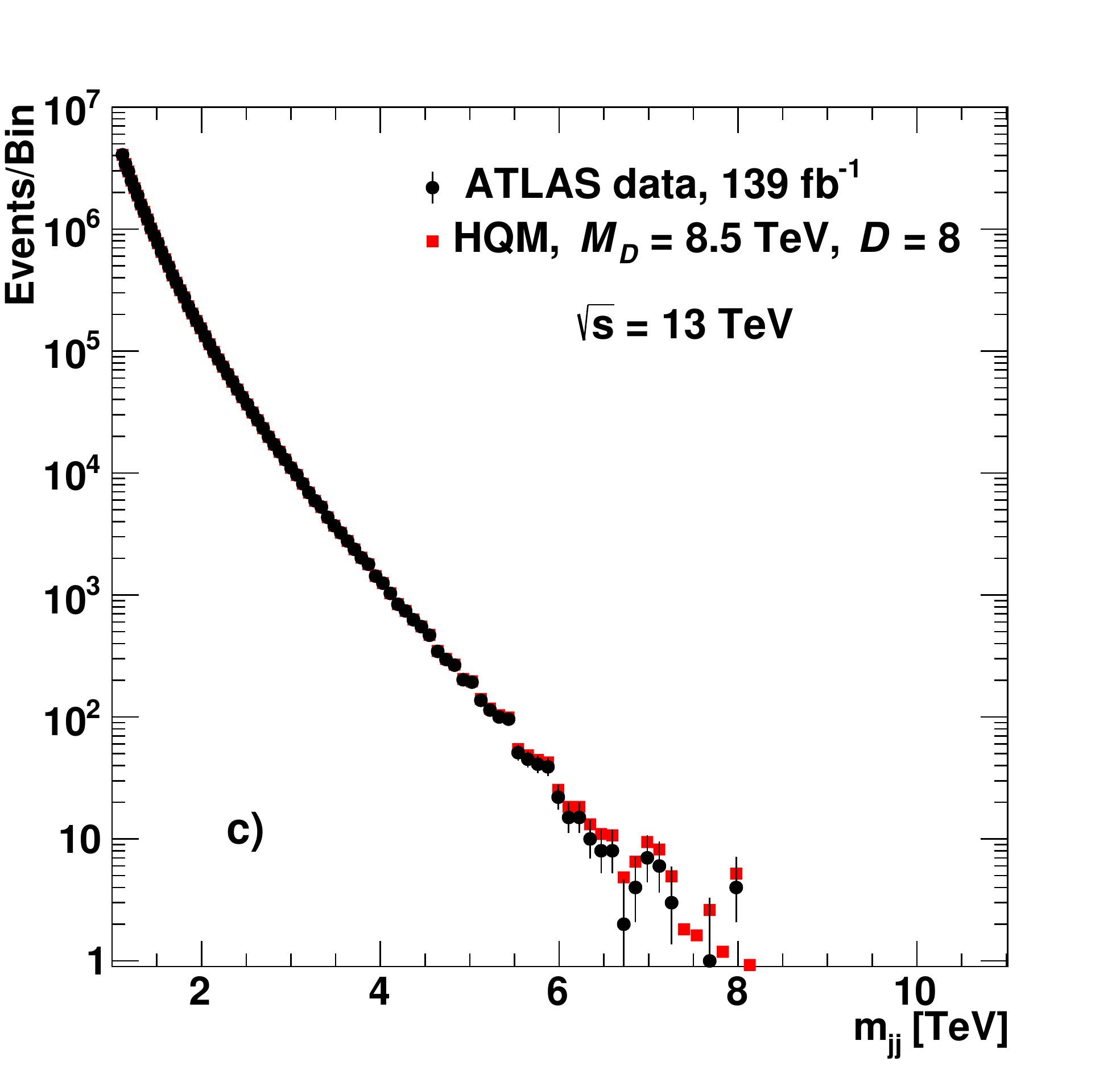}
\includegraphics[width=7cm]{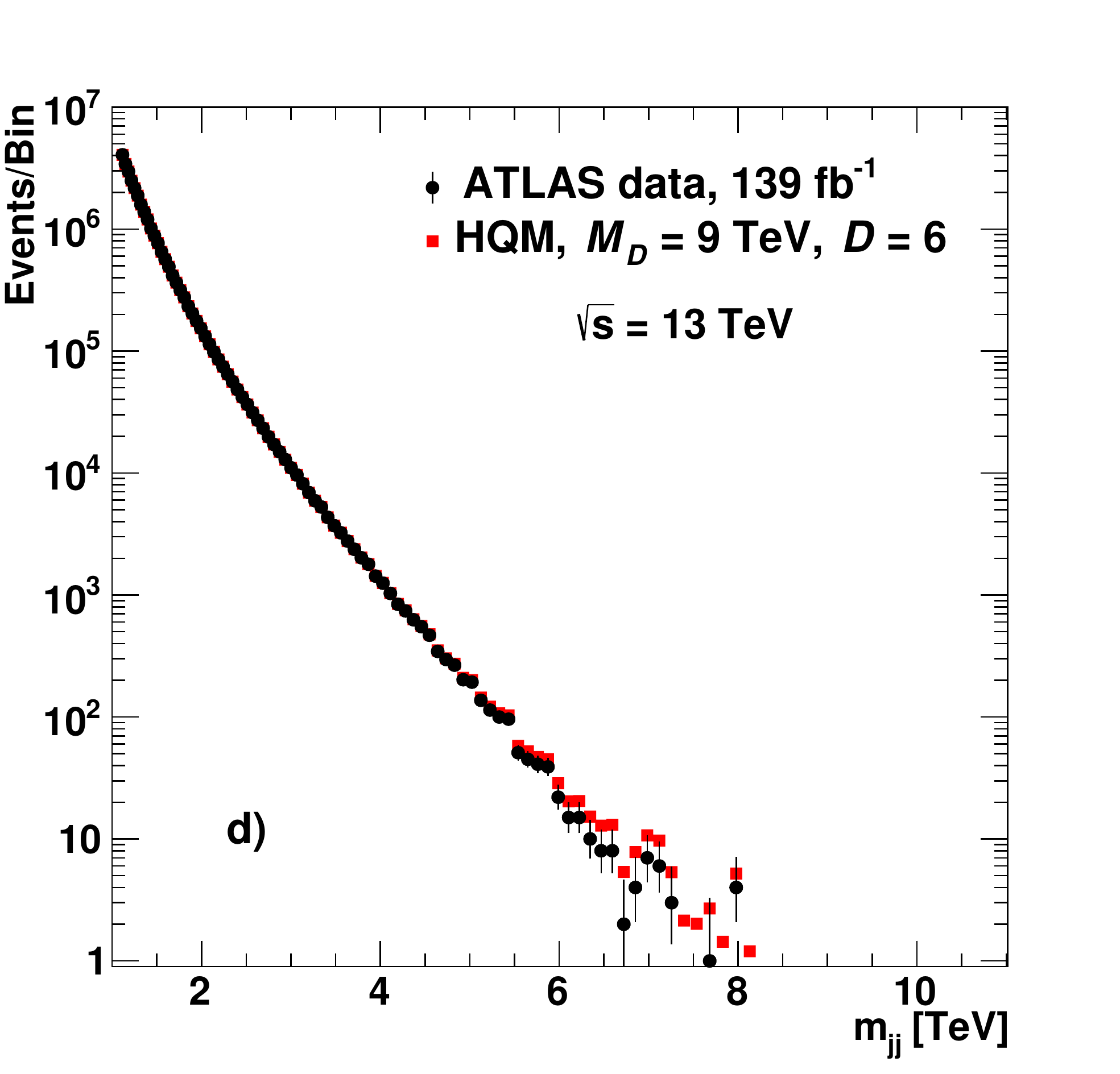}
\caption{\label{fig07}Black hole dijet mass distributions scaled by
cross section to the ATLAS data luminosity and added on top of the ATLAS
dijet mass m$_\mathrm{jj}$ spectrum measured at a center of mass energy
of 13~TeV and a luminosity of 139~fb$^{-1}$~\cite{Aad:2019hjw}.   
The results are shown for the quantum black hole (QBH) model and horizon
quantum mechanics (HQM) model for various values of the Planck scale \MD
and the total number of extra space-time dimensions $D$.}
\end{figure*}

For the HQM model, dijet distributions are shown in
Fig.~\ref{fig07}b),
Fig.~\ref{fig07}c), and
Fig.~\ref{fig07}d) for
($\MD = 8$~TeV,   $D=10$),
($\MD = 8.5$~TeV, $D=8$), and
($\MD = 9$~TeV,   $D=6$), respectively.
Although ATLAS and CMS have not set limits on the HQM model they have
eliminated a wide variety of resonances in the dijet mass spectrum
from trigger turn on to about 8~TeV.
Thus HQM black hole production resulting in sizable deviations from
the smoothly falling dijet mass distribution are not allowed.
The values of \MD in the figures have been chosen high enough to not
result in a clear enhancement in the dijet mass distribution that ATLAS
and CMS have not seen. 
On the other hand, if the \MD values are chosen higher the number of
events becomes insignificant for masses above the ATLAS and CMS data
points. 
It would thus be extremely difficult to observe black holes in the HQM
model in the current dijet invariant mass spectrum.

\section{Discovery potential in the dijet mass distribution}

In order to predict the discovery potential for observing quantum black
holes, we take into consideration both the number of events above
background and the significance of the signal.
For the significance, we use the asymptotic approximation without
background uncertainty (see, for example Ref.~\cite{Cowan:2010js}).
The formula comes from using the asymptotic formulae for the
distributions of profile likelihood test statistics.  

\begin{equation}
\sigma = \sqrt{2\left[\left(s+b\right)\ln\left(1+\frac{s}{b}\right)
    -s\right]}\, ,  
\label{eq4}
\end{equation}

\noindent
where $s$ is the number of signal events above background and $b$ is the
number of background events excluding signal events.
The signal events are generated with \qbh and the ATLAS background model
is taken as the background. 
We understand that Eq.~(\ref{eq4}) is an approximation based on a
cut-and-count approach, and that one should really include background
uncertainties.
However, such an analysis is beyond the scope of this work, and is
unlikely to change the qualitative findings.

We consider a significant observation to be greater than $5\sigma$.  
Using a cut-and-count method, significance is calculated by counting
events above \Mth. 
While this is natural for the QBH model, it is perhaps not so meaningful
for the HQM model since many of the events have $M < \MD$. 
For the sake of comparison, we consider two approaches to calculating
the significance for the HQM model. 
The first is the usual definition, where we consider \MD as a cutoff. 
In this method \MD values beyond $\sqrt{s}$ can not be probed.
In the second method, we consider all black hole events and count the
background from the least massive signal event. 
We understand that the latter method would be extremely difficult, and
probably not even desirable, to realize in an experiment's analysis,
but it might be more indicative of a shape-fit procedure that might
likely be used.  

The event count and significance are presented in
Fig.~\ref{fig08} and Fig.~\ref{fig09}, respectively. 
While counting HQM model events over the entire mass range gives the
greater number of events, the method of counting HQM model events only
above \MD give better significances.  
This could have been anticipated given the large number of background 
events at low dijet masses.
Using either approach to calculating the significance, the discovery
potential at allowed values \MD is less for the HQM model than the
QBH model.
Since the ATLAS background that we are using does not extend beyond
8.1~TeV, and because of the simple significance formula Eq.~(\ref{eq4}),
the significance curves in Fig.~\ref{fig09} end at $\MD = 8$~TeV.

\begin{figure}[p]
\centering
\includegraphics[width=0.82\columnwidth]{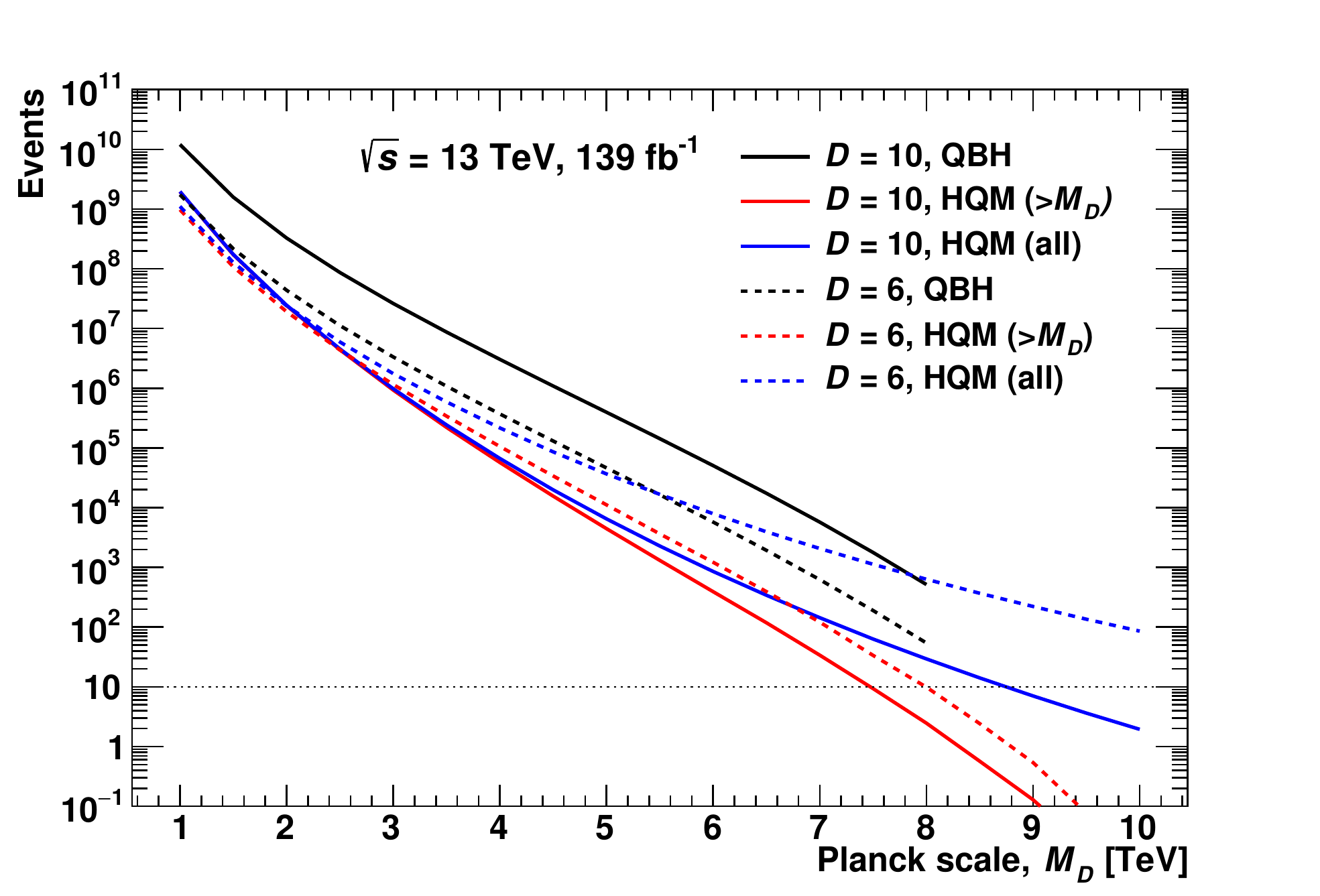}
\caption{\label{fig08}Predicted number of black hole events versus
  Planck scale \MD for a center of mass energy of 13~TeV and
  luminosity of 139~fb$^{-1}$ when selecting events at the parton
  level according to the same criteria as the search in
  Ref.~\cite{Aad:2019hjw}.  
The solid curves are for total space-time dimension $D=10$ and the
dashed curves for $D=6$.}
\end{figure}

\begin{figure}[p]
\centering
\includegraphics[width=0.82\columnwidth]{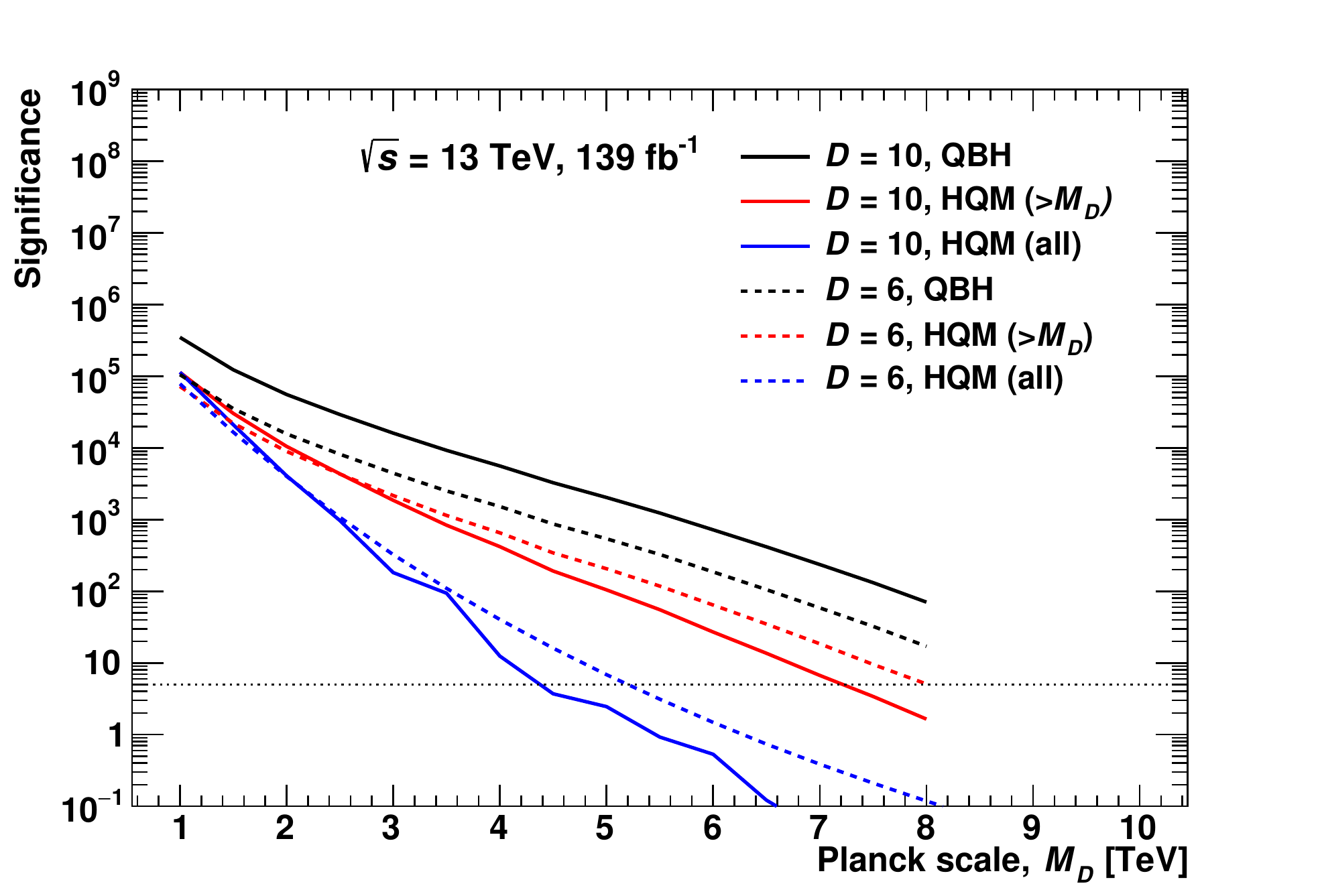}
\caption{\label{fig09}Significance of a black hole observation above
  dijet background versus Planck scale \MD for a center of mass energy
  of 13~TeV and luminosity of 139~fb$^{-1}$.
Event were selected at the parton level according to the same criteria
as the search in Ref.~\cite{Aad:2019hjw}. 
The solid curves are for total space-time dimension $D=10$ and the
dashed curves for $D=6$.}
\end{figure}

Using the $M > \MD$ counting method and by noting the minimum \MD value
given by the ten event and $5\sigma$ criteria, we assess the possibility
of detecting HQM black holes in ATLAS and CMS. 
For $D = 10$, the number of signal events is greater than ten for
$\MD \lesssim 7.5$~TeV.
The corresponding significance is greater than $5\sigma$ for $\MD
\lesssim 7.4$~TeV, and this sets the upper limit on \MD to observe
black holes in the HQM model.
For the $D = 6$ case, greater than ten events occurs when $\MD \lesssim
8.0$~TeV and the significance is greater than $5\sigma$ at $\MD
\lesssim 8.0$~TeV.
However, with only one background event, the significance as defined in
Eq.~(\ref{eq4}) slightly overestimates the true significance.
In any case, the lower limit on \MD from the CMS
experiment~\cite{Sirunyan:2017jix} for $D = 6$ is 9.9~TeV at the 95\%
confidence level, thus eliminating the HQM model for $D=6$. 

Given the increase in luminosity and $\sqrt{s}$ in subsequent LHC runs,
these discovery potentials stand to increase somewhat.
With this thought in mind, we make some predictions at $\sqrt{s} =
13$~TeV on the luminosity required at a given \MD for a meaningful
discovery. 
We assume that the number of background events, based on the background
model from
Ref.~\cite{Aad:2019hjw}, scales linearly with luminosity. 
When calculating the significance using $M > \MD$ as a cutoff in the
cut-and-count method, we have made the additional assumption that
event-count is the limiting factor for $\MD > 8$~TeV as this is the
highest dijet mass at which the ATLAS background estimate is given. 
The results are shown in Fig.~\ref{fig10} where we only consider
luminosities above 139~fb$^{-1}$.
The luminosity axis of the plot extends out to 4000~fb$^{-1}$, inspired
by the design integrated luminosity of the High-Luminosity Large Hadron
Collider.  
It is seen that the increase in probing \MD with a reasonable increase 
in luminosity is not very significant, indicating that we are close to
exhausting the search for black holes in both QBH and HQM models using 
the dijet mass distribution at $\sqrt{s} = 13$~TeV.
Although we have used a very simplistic approach to estimating the
discovery potential, this conclusion is unlikely to change with a more
robust estimate.

\begin{figure}[tbp]
\centering
\includegraphics[width=0.9\columnwidth]{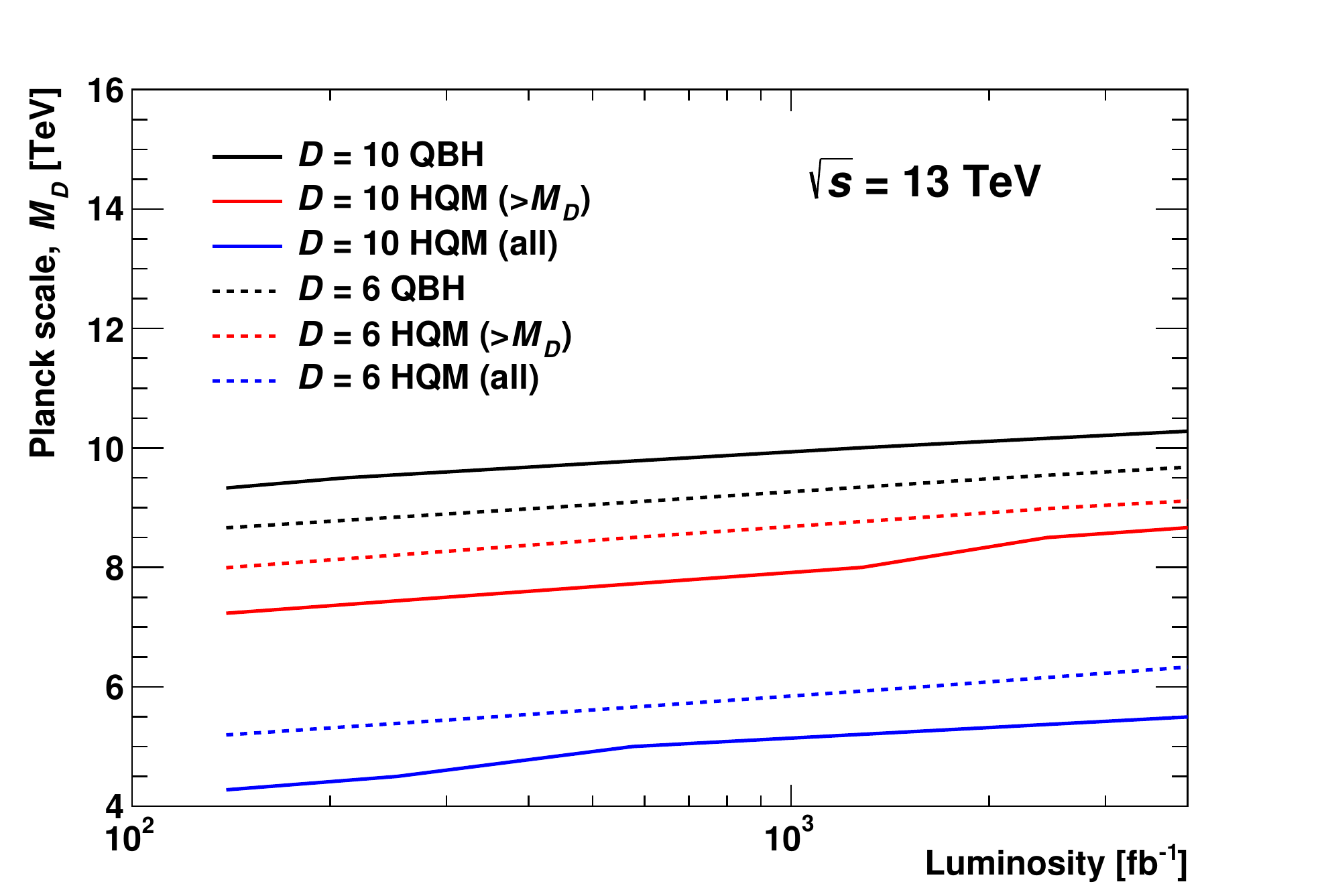}
\caption{\label{fig10}Minimum luminosity required to produced at least
ten signal events and a significance of $5\sigma$ at a center of
mass energy of 13~TeV.
The solid curves are for total space-time dimension $D=10$ and the
dashed curves for $D=6$.}
\end{figure}


\section{Conclusions}

Microscopic black hole formation as predicted by HQM was implemented in
the \qbh MC event generator to investigate the impact on possible black
hole production at the LHC.  
The inclusion of the HQM model serves to decrease the total black hole
cross section for small \MD, but the new model is not restricted by a
threshold mass requirement. 
Therefore, HQM predicts black holes may be produced at $\MD \sim
\sqrt{s}$.  
The HQM model is also highly dependent on dimensionality and predicts
that a greater number of events may be produced with a smaller $D$. 
The $M$ distribution is also greatly affected by HQM with a much wider 
spread of black hole masses.  
In other words, there is no resonance structure in the HQM model. 
This wide $M$ distribution converges to a constant shape for large
\MD, which can be considered to be one of the defining features of
the HQM model.

The predicted signal in the dijet mass distribution along with the ATLAS
run-2 background model were used to estimate the number of signal events
and significance. 
Observations of quantum black holes governed by HQM were predicted to be
limited to $\MD \lesssim 8.0$~TeV for $D = 6$ and $\MD \lesssim 7.2$~TeV
for $D = 10$.  

Given the small potential for observation of HQM black holes in the
dijet mass distribution, a discovery in the invariant mass variable
is unlikely.  
Alternatively, an angular search may be performed to distinguish an
enhancement of events due to black hole production above QCD 
background~\cite{Aad:2011aj,ATLAS:2012pu,ATLAS:2015nsi,Aaboud:2017yvp,Sirunyan:2017ygf}. 
The HQM model does not yet predict any modification to the usual decays
in the QBH model; there is no difference between the two models in terms of
the shape of angular distributions. 

An example angular search could be in the variable $\chi$ defined as

\begin{equation}
\chi = e^{|y_1-y_2|}\, , 
\label{eq5}
\end{equation}

\noindent
where $y_1$ and $y_2$ are the rapidities of the two jets.
QCD $t$-channel scattering constituting the background is approximately
constant in $\chi$, while $s$-channel resonances tend to be enhanced at
low $\chi$.
Because of this, an angular search could help uncover the wide
$s$-channel mass enhancement that is predicted by HQM. 
Since the predictions of an angular search are highly dependent on the
analysis and detector details, we leave it to the ATLAS and CMS
collaborations to perform such a search. 

Some of the above results were first mentioned in
Ref.~\cite{Arsene:2016kvf}.  
Unfortunately, that paper could only make use of ATLAS and CMS results
from about 20~fb$^{-1}$ of data at a center of mass energy of 8~TeV.
We view our analysis as more comprehensive, benefiting from using
recently available experimental data distributions, and up to date. 

Lastly, although the HQM model has been used, we do not believe the
qualitative results presented here depend specifically on the formula
presented in Ref.~\cite{Casadio:2015jha}; similar results would be
obtained for any non-step-like threshold mass production of black
holes such as those presented in Ref.~\cite{Mureika:2011hg}.

We have implemented and studied a benchmark model and some of the
quantitative results are model dependent.
It is not our intent to prove or disprove a particular model but to
point out the need for alternative search strategies for quantum black
holes such as a dijet angular analysis.

\begin{acknowledgments}

This work was supported in part by the Natural Sciences and Engineering
Research Council of Canada.
\end{acknowledgments}

\appendix
\section{Monte Carlo event generation}
\label{appA}

In order to visualise how the HQM probability \PBH varies with $D$, \MD,
and $M$, we computed the integral in Eq.~(\ref{eq2}) explicitly using 
numerical integration.  
As shown in Fig.~\ref{fig01}, good accuracy was achieved with the
use of Simpson's method and an adequate large number of subdivisions.

A more elegant means of producing the appropriate \PBH factor
can be performed by MC integration.
As a check, we have also produced the curves in Fig.~\ref{fig01} using
MC sampling.  
By integrating and inverting the \PH distribution, random 
values of the horizon radius $\rH$ can be sampled using a uniform 
distribution of random numbers. 
Since \PH is a probability density function, using random $\rH$ values
to calculate \PS for a large number of samples effectively computes the
expected value for \PS (or equivalently, \PBH).
For completeness, we present this calculation.

We begin from Eq~3.7 in Ref.~\cite{Casadio:2015jha}:

\begin{eqnarray}
\PH(\rH) & = & a^\frac{d}{d-2} \frac{2(d-2)}{\Gamma\left( s, 1\right)}
\Theta(\rH - R_d)\nonumber\\
& & \times \exp \left( -a^2 \rH^{2(d-2)} \right) \rH^{d-1}\, ,
\label{eq6}
\end{eqnarray}

\noindent
where in this appendix we use the notion of Ref.~\cite{Casadio:2015jha}
except we take the total number of spatial dimensions to be $d$.
We have define $a=(d-2)/(2m)$ and $s = d/[2(d-2)]$, and used $\Delta =
m$ as in Ref~\cite{Casadio:2015jha}; $\Gamma(s,x) = \int_x^\infty
t^{s-1} e^{-t} dt$ is the upper incomplete Gamma function. 
In addition, we are using Planck units since we are only interested in
lengths and masses relative to \MD.

By taking the Heaviside step function to be one, the indefinite integral
can be computed:

\begin{equation}
\int \PH(\rH) d\rH = - \frac{\Gamma\left( s, a^2
  \rH^{2(d-2)}\right)}{\Gamma\left(s, 1\right)}\, .  
\label{eq7}
\end{equation}

\noindent
Substituting a lower limit of $R_d=[2m/(d-2)]^{1/(d-2)}$ and upper
limit of $\rH$, to allow calculation of the cumulative distribution
function, gives 

\begin{equation}
\mathrm{CDF}[\PH(\rH)] = 1 - \frac{\Gamma\left( s, a^2
  \rH^{2(d-2)}\right)}{\Gamma\left(s, 1\right)}\, .
\label{eq8}
\end{equation}

If we generate a uniform random real number $u$ in the interval $(0,1)$
[or $1-u$ in the interval $(1,0)$] and set it equal to
Eq.~(\ref{eq8}), we can solve for $\rH$ by inverting the incomplete
Gamma function with respect to its second parameter:

\begin{equation}
\rH = R_d \left[ Q^{-1}\left( s, Q(s,1) u \right)
\right]^\frac{1}{2(d-2)}\, .
\label{eq9}
\end{equation}

\noindent
Note that $Q^{-1}(s,Q(s,x)) = x$, where $Q^{-1}$ is the inverse of the
regularized upper incomplete Gamma function $Q(s,x) =
\Gamma(s,x)/\Gamma(s)$. 
There are numerical methods to optimise this inversion.

Upon randomly sampling the horizon radii from Eq.~(\ref{eq9}), we return
values of $\PS(r<\rH)$ as given by Eq.~3.5 in Ref.~\cite{Casadio:2015jha}:   

\begin{equation}
\PS(r<\rH) =
\frac{\gamma\left( \frac{d}{2},m^2\rH^2\right)}{\Gamma\left(
\frac{d}{2}\right)}\, ,
\label{eq10}
\end{equation}

\noindent
where $\gamma(s,x) = \int_0^x t^{s-1} e^{-t} dt$ is the lower incomplete
Gamma function.

The above random horizon generation can simply be looped over with an
average of all \PS values giving an approximate value for \PBH. 
We easily recreate the same probability curves as in Fig.~\ref{fig01}
which used Simpson's method.

Both the MC method and Simpson's method for calculating \PBH have been
implemented in \qbh.
Despite both methods producing the same results, there are technical
pros and cons of each method.
The MC HQM calculation just presented is the default method.

One additional technicality should be mentioned.
Since black hole production in the HQM model allows for $M$ less than
\MD there is no lower-mass cutoff in the generator.
Instead, the \PBH curve imposes its own smooth limit as it
becomes arbitrarily small.
To sample $M$ via a power transformation of the cross section used to
increase efficiency, we choose an arbitrary minimum of 100~GeV since in 
practise it is exceedingly rare to generate an event with $M$ this
low. 
For example, selecting a 200~GeV minimum has a negligible impact on the
results. 

We point out that our curves of \PBH are identical to the
corresponding figure in Ref.~\cite{Arsene:2016kvf} within our ability to
read values from their figure.
Equation~(7) in Ref.~\cite{Arsene:2016kvf} disagrees with Eq.~(3.8)
Ref.~\cite{Casadio:2015jha}, although the later cites the former.
We believe Eq.~(7) in Ref.~\cite{Arsene:2016kvf} has the inverse power
of $(m_d/m)$ and a normalization difference of $(D-2)^2$.
If the formula in the paper was actually use to generate the plot, the
curves continue to increase above unity with increasing mass and do not
represent probability distributions.

\bibliographystyle{apsrev4-2}
\bibliography{hqm}
\end{document}